\magnification\magstep1 
\baselineskip14pt 
\vsize23.5truecm

\font\smallrm=cmr8

\def\hopp{\medskip\noindent}
\def\hop{\smallskip\noindent} 
\def\E{{\rm E}} 
\def\Var{{\rm Var}}
 
\def\N{{\rm N}} 
\def\Pr{{\rm Pr}} 
\def\EE{{\cal E}}
 
\def\risk{{\rm risk}} 
 
\def\dell{\partial} 
\def\arr{\rightarrow}
\def\hatt{\widehat} 
 
\def\sumin{\sum_{i=1}^n}

\def\eps{\varepsilon} 
\def\half{\hbox{$1\over2$}}
 
\def\section{\bigskip}
\def\subsection{\medskip}

\def\true{{\rm true}} 
\def\full{{\rm full}}

\def\FIC{{\rm FIC}}

\def\const{{\rm const.}} 
\def\rootn{\sqrt{n}}

\def\midd{\,|\,}  
\def\tr{{\rm t}}
\def\Tr{{\rm Tr}}

\def\dellone{\hbox{$\dell\mu\over \dell\theta$}}

\def\delltwo{\hbox{$\dell\mu\over \dell\gamma$}}
\def\diff{\omega}

\def\fermat#1{\setbox0=\vtop{\hsize4.00pc
        \smallrm\raggedright\noindent\baselineskip9pt
        \rightskip=0.5pc plus 1.5pc #1}\leavevmode
        \vadjust{\dimen0=\dp0
        \kern-\ht0\hbox{\kern-4.00pc\box0}\kern-\dimen0}}

\centerline{\bf Rejoinder to the discussants}
\centerline{\bf Nils Lid Hjort and Gerda Claeskens}

\smallskip
\centerline{\sl -- August 2003 --}

\hopp
We are honoured to have our work read and discussed 
at such a thorough level by several experts.
Words of appreciation and encouragement are gratefully
received, while the many supplementary comments, 
thoughtful reminders, new perspectives and additional themes raised 
are warmly welcomed and deeply appreciated. 
Our thanks go also to JASA Editor Francisco Samaniego
and his editorial helpers for organising this discussion.

Space does not allow us answering all of the many
worthwhile points raised by our discussants,
but in the following we make an attempt to respond
to what we perceive of as being the major issues. 
Our responses are organised by themes rather than by discussants. 
We shall refer to our two articles as `the FMA paper' and `the FIC paper'. 

\section 
\centerline{\bf 1. The local neighbourhood framework}

\hop
In our articles we chose to work inside a broad and general
parametric framework, 
which in the regression case corresponds to our using say
$$f_{i,\true}(y)=f(y\midd x_i,\sigma_0,\beta_0,
   \gamma_0+\delta/\rootn)\,; \eqno(1.1)$$ 
see Section 2 in the FMA and Section 2 in the FIC paper.  
This draws partial criticism from Raftery and Zheng,
who question its realism, as well as from Ishwaran and Rao,
who argue that it does not yield a good framework
for subset regression problems. 

The local neighbourhood framework (1.1) allows one to extend
familiar standard i.i.d.~and regression models 
(corresponding to having $\delta$ fixed at zero) 
in several parametric directions
(corresponding to $\delta_1,\ldots,\delta_q$ allowed to be non-zero, 
for different envisaged departures from the start model),
as exemplified in our papers. This may in particular be utilised 
for robustness purposes and sensitivity analyses, and leads 
to a fruitful theory for model averaging and focussed model selection
criteria, as we have demonstrated. 

In their Section 4, Raftery and Zheng mention two pro (1.1) arguments,
before presenting their scruples. The main argument for 
working inside (1.1) is however that it leads to 
natural, general and precise limit distribution results, 
with consequent approximations for mean squared errors and the like;
the key is that variances and squared modelling biases
become exchangeable currencies, both of size $1/n$. 
For classes of estimators $\hatt\mu$ of $\mu(\theta,\gamma)$, 
including the submodel estimators 
$\hatt\mu_S=\mu(\hatt\theta_S,\hatt\gamma_S,\gamma_{0,S^c})$, 
we have 
$$\E_{\theta,\gamma}\{\hatt\mu-\mu(\theta,\gamma)\}^2
  =n^{-1}\rho_1(\theta,\rootn(\gamma-\gamma_0))
   +n^{-3/2}\rho_2(\theta,\rootn(\gamma-\gamma_0))
   +\cdots, \eqno(1.2)$$
for example, under regularity conditions. Such expansions,
written out here without the $\delta$ that Raftery and Zheng
appear to dislike, would typically be valid uniformly
over $\|\gamma-\gamma_0\|\le\const/\rootn$ balls. 
We view (1.2) type results as a good reason 
for developing and presenting theory in terms of 
$\delta=\rootn(\gamma-\gamma_0)$, i.e.~using (1.1). 
Our papers have (in particular) provided formulae 
for $\rho_1(\theta,\delta)$ here, the limiting risk
for $\hatt\mu$, while expressions for $\rho_2(\theta,\delta)$ 
are harder to get hold of; see our response below 
to Tsai's comments. We have also noted, in FIC's Section 5.5, 
that approximations coming from using the leading term 
in (1.2) type expansions hold with exactness for finite $n$,
for submodel estimators of means in linear regression. 

Thus Raftery and Zheng interpret us a little bit too literally, 
at the end of their Section 4; 
as statisticians we do not believe that our 
model parameter $\gamma$ changes value when our 
data set passes from $n=100$ to $n=101$, but we do believe 
that limit theorems based on the (1.1) framework
provides lucent understanding and useful approximations
for the given $n$. This comment also applies to our 
BMA investigations (FMA's Section 9), where priors and
posteriors for $(\theta,\gamma)$ are transformed to 
priors and posteriors for $(\theta,\delta)$. (A too literal
belief in sample-size dependent parameters would clash
with Kolmogorov consistency and other requirements for
natural statistical models; see McCullagh (2002) 
and the ensuing discussion.)

A further strand of arguments supporting the view that 
many questions find their most natural solutions inside  
$\gamma-\gamma_0=O(1/\rootn)$ frameworks is related 
to what we termed `tolerance radii' in FMA's Section 10.5. 
How much quadraticity, or variance heteroscedasticity, 
can the normal regression model tolerate, in the sense
that the simpler methods based on standard assumptions
still give better results than the more cumbersome
ones based on the bigger models? 
How much autocorrelation can typical i.i.d.-based methods take? 
Such questions are nicely answered using the sample-size dependent 
magnifying glass $\delta=\rootn(\gamma-\gamma_0)$, as touched on 
in Section 10.5 of the FMA paper. Consider for example
the $(\beta_0,\sigma,\beta_1,\lambda)$ model of FIC's Section 4.1.
The simple i.i.d.~model $Y_i\sim\N(\beta_0,\sigma^2)$ 
can tolerate the presence of a regression coefficient $\beta_1$ 
and a skewness parameter $\lambda$, as long as 
$$\rootn|\diff_1\beta_1+\diff_2(\lambda-1)|\le
   (k_{n,1}\diff_1^2+k_{n,2}\diff_2^2)^{1/2}. $$
This first-order asymptotic answer rests on a framework
like (1.1), and depends also on the focus parameter under study
via $(\diff_1,\diff_2)$; see FIC's Section 4.1 for examples. 
Inside the ellipse $k_{n,1}\beta_1^2+k_{n,2}(\lambda-1)^2\le 1/n$,
{\it all} estimands will be better estimated using the simple
$\N(\beta_0,\sigma^2)$ model than using the the formally correct
four-parameter model. 

Another benefit of our methodology, and the (1.1) type framework, 
is the ability to compare model selection and model average strategies 
in a unified way, across situations, so to speak; 
there is a well-defined limit experiment, characterised 
by deterministic quantities $\tau_0$ and $K$,
the vector $\diff$ which depends on the focus parameter, 
plus an unknown $\delta$ for which one observes 
$D\sim\N_q(\delta,K)$. Inference is then sought for 
$\psi=\diff^\tr\delta$. Thus lessons learned for e.g.~Poisson
regression models can be carried over to e.g.~logistic regressions.
This is also in the LeCam spirit of asymptotic equivalence of statistical
experiments; see e.g.~van der Vaart (1998) and Brown (2000)
for general discussion. 

Note that $\delta$ can be big in size, in (1.1) and (1.2),
so reading our papers as saying that we only care about 
$\gamma$ being close to $\gamma_0$ is not correct. Also,
when $\gamma$ happens to be far away from $\gamma_0$,
this will be picked up by the data via 
$\hatt\delta_\full=\rootn(\hatt\gamma_\full-\gamma_0)$,
and most sensible model averaging schemes, including 
FIC and weighted FIC, will give weights close to 1 for 
the wide model. 
As e.g.~Johnson hints at, other techniques might be 
needed to better assess behaviour and properties of 
model average methods in clearly non-contiguous situations,
i.e.~when $\gamma$ is far from $\gamma_0$.  

We have seen that the $O(1/\rootn)$ framework is canonical 
inside general parametric models with independence. 
This has to do with information increasing linearly 
with sample size divided by model complexity and 
variances being proportional to inverse information.
Ishwaran and Rao mention Breiman's bagging, which indeed 
may be viewed in terms of model averaging. Some of the calculations 
in Section 2 of B\"uhlmann and Yu (2002) may be seen as 
special cases of our general FMA theory. Their paper shows
that when averaging takes place over a large number of stumps,
then (su)bagging is best analysed inside a $O(1/n^{1/3})$ framework. 
A similar comment applies to some of the goodness-of-fit tests
of Claeskens and Hjort (2003), where large classes of 
alternative models are being searched through. 

\section
\centerline{\bf 2. Two uses of regression models}

\hop
Regression analyses have different aims on different
occasions, and even the same data set may be analysed 
with different goals in mind. We have primarily taken
the view that what matters most is quality of predictors
and precision of focus estimators.  
Ishwaran and Rao, in contrast, equate the subset selection 
regression problem with finding the exact subset of non-zero 
elements among a vector of coefficients $(\beta_1,\ldots,\beta_K)^\tr$. 
As a result they partly criticise our methods for not
being optimal for a task they were not set out to perform. 

For many applications one would not care much if 
say $\beta_7=0.01$ rather than being exactly zero,
and the additional estimation noise caused by including
$\hatt\beta_7$ in the predictor formulae might worsen 
rather than enhance the precision. Ishwaran and Rao appear
to say that it is the duty of any subset selection method
to strive for inclusion also of such a small $\beta_7$. 

There are of course situations where detecting the
non-zero-ness of certain parameters is the main goal
of an analysis. This could be a $\beta_j$ coefficient
in a linear regression setting, for example. 
Our theory works for such a focus parameter too, 
since it may be expressed as 
$\beta_j=\E(Y\midd x+e_j,u)-\E(Y\midd x,u)$,
with $e_j$ the $j$th unit vector. In an effort towards making
the world slightly less unfair, Hjort (1994a) collected and 
analysed data from world championships in sprint speedskating,
attention focussed on the average difference $d$ between
500 m results reached using last inner track vs.~last outer track. 
The analysis essentially employed a bivariate mixed effects regression model 
with seven parameters per championship (and it was necessary 
to fit the full seven-parameter model to make inference for $d$). 
Only one parameter mattered to the delegates from 37 nations 
at the 1994 general assembly of the International Skating Union. 
They had to assess the potential non-zero-ness of $d$
and its implications, and actually needed to vote for or against 
the significance of the point estimate 
(which was $\hatt d=0.06$ seconds).  
The Olympic rules for sprint speedskating
were in fact changed as a consequence of the statistical analysis;
from Nagano 1998 onwards the athletes are forced to skate the 500 m distance 
twice. See also Hjort and Rosa (1999). This is an example of a sharply
defined focus parameter where tools of FMA and FIC might be used.

We disagree with the way Ishwaran and Rao interpret 
the scope of our machinery for subset selection problems
in regression, at the end of their Section 1. The statistician
is at the outset required to classify some parameters 
(say $\beta_j$s) as `protected' and other parameters 
(say $\gamma_j$s) as `uncertain'; our methods are then
geared towards finding the best subsets of $\gamma_j$s 
to include, or to be averaged over. Our methods are certainly
not `restricted to coefficients known to be zero', as 
Ishwaran and Rao charge in their point (a). 
First, the methods are well-defined 
and can be applied regardless of the sizes of the $\gamma_j$s.
This is also a reply to a comment by Raftery and Zheng,
that our (1.1) is `required by FMA'; FMA methods give
algorithms that may be put to work regardless of (1.1) type assumptions.
Second, even though the mathematical results we have provided
about various methods have utilised the 
$\gamma_j=\gamma_{0,j}+\delta_j/\rootn$ framework, 
the $\delta_j$s may be big in size, as also commented on above. 
Third, most sensible selection methods or averaging methods 
will pick out the widest model, in cases where the $\gamma_j$s 
really are far from zero. 

Regarding their point (b) (end of Section 1), it is fair to
say that statistical modelling is and remains an art 
demanding skill and experience for its perfect execution,
even with the advent of further tools for automatisation
and diagnostics. 
The previous argument indicates that it may be rather harmless 
if a statistician labels a parameter a `$\gamma$'
when it should rather have been a `$\beta$' (provided the
selection or averaging scheme is among the decently robust 
ones, with low max risk, see FMA's Section 7); this also serves
to counter their point (b). A similar comment applies to
Raftery and Zheng's reservations (Section 4), having 
to do with situations where `the coefficients for some nuisance variables
are substantial, and those for others are small'. 
In such cases the crafty modeller should take this on board,
redesigning nuisance variable coefficients as protected. 
 
In their Sections 1 and 2, Ishwaran and Rao argue that in most regression
setups, the $\gamma_0$ associated with uncertain (or non-protected) 
variables must be zero. This is fine, is not surprising, and does not 
contradict our machinery or methodology.  
Our theory does allow $\gamma_0\not=0$ too,
but this would here correspond to known trends, 
which may be removed from the regression equation.
We note that the FIC paper has several examples where
the canonical $\gamma_0$ is non-zero. 

\section 
\centerline{\bf 3. Estimating model order}

\hop
In some settings there is a natural order of complexity 
among candidate models, as with e.g.~polynomial regression.
Ishwaran and Rao (Section 3) study the problem of estimating the actual
underlying order of the true model, that is, the unknown
number $k_0$ where the coefficient vector is 
$(\beta_1,\ldots,\beta_{k_0},0,\ldots,0)$ and 
the first $k_0$ are strictly non-zero. As mentioned above, 
in many situations estimating $k_0$ might not be a vital issue.
Their Theorem 1 contrasts backwards and forwards selection schemes
under some conditions, and is of interest. We believe 
their theorem should and can be extended to more general
settings, however. 

We do think the assumption about finiteness of fourth moments
may be softened, although this is not crucial. 
The primary problem we see is their assumption that  
$\Sigma_n=n^{-1}X^\tr X=n^{-1}\sumin x_ix_i^\tr$ 
must equal the identity matrix $I$; this appears too restrictive. 
One may transform a regression model to achieve such orthogonality,
but this would typically inflict a different ordering of 
new coefficients, losing the original motivation of nested-ness.
This makes it difficult to keep track of the original $k_0$.
To avoid the problem one may keep the original model, 
and consequently keep the $k_0$ as defined by the untransformed
$\beta$ vector, but accept the weaker assumption that 
$\Sigma_n$ tends to a general positive definite $Q$.  

Under weak Lindeberg type conditions,
see e.g.~Hjort and Pollard (1994) 
(in particular, it does not appear necessary to assume 
finite fourth moments), we then have 
$\rootn(\hatt\beta-\beta)\arr_d\N_K(0,\sigma^2Q)$,
with consequent $\rootn(\hatt\beta_k-\beta_k)\arr_d\N(0,\sigma_k^2)$,
say, where $\sigma_k=\sigma(Q_{k,k})^{1/2}$. 
Thus there is simultaneous convergence 
$Z_{k,n}=\rootn\hatt\beta_k/\hatt\sigma_k\arr_d Z_k$,
say, for $k\ge k_0+1$, where these are standard normals 
with correlations inherited from $Q$. Also, $|Z_{k,n}|$ flees
to infinity in probability for $k\le k_0$. Defining the backwards
and forwards model order estimates as in Ishwaran and Rao,
one may now show that $\hatt k_B\arr_d k_B$
and $\hatt k_F\arr_d k_F$, where 
$$\Pr\{k_B=k\}=\cases{0 &for $k\le k_0-1$, \cr
   \Pr\{Z_{k_0+1}\in J_{k_0+1},\ldots,Z_K\in J_K\} 
      &for $k=k_0$, \cr
   \Pr\{Z_k\notin J_k,Z_{k+1}\in J_{k+1},\ldots,Z_K\in J_K\} 
      &for $k\ge k_0+1$ \cr}$$
and 
$$\Pr\{k_F=k\}=\cases{0 &for $k\le k_0-1$, \cr
   \Pr\{Z_{k_0+1}\in J_{k_0+1}\} 
      &for $k=k_0$, \cr
   \Pr\{Z_{k+1}\in J_{k+1},Z_{k_0+1}\notin J_{k_0+1},\ldots,Z_k\notin J_k\} 
      &for $k\ge k_0+1$. \cr}$$
Here $J_k=(-z_{\alpha_k/2},z_{\alpha_k/2})$ is the acceptance
interval for $Z_{k,n}$, with limit probability $1-\alpha_k$. 
Ishwaran and Rao's Theorem 1 corresponds to the case of 
a diagonal $Q$ matrix, where the $Z_k$s become independent. 

In practice the $|Z_{k,n}|$s for $k\le k_0$ have not quite 
had time to flee to infinity, for finite $n$, as $Z_{k,n}$
has mean value about $\rootn\beta_k/\sigma_k$. The approximations
afforded by the limit theorem above are easily too crude,
particularly when $\beta_k$s are small. It is again natural
to use the local neighbourhood parametrisation, with say
$\beta_k=\delta_k/\rootn$. The limit distributions for 
$\hatt k_B$ and $\hatt k_F$ may be derived. There will 
in particular be positive probabilities for values $k\le k_0-1$.
One finds in fact 
$$\Pr\{k_B=k\}=\Pr\Bigl\{{\delta_k\over \sigma_k}+Z_k\notin J_k,
   {\delta_{k+1}\over \sigma_{k+1}}+Z_{k+1}\in J_{k+1},\ldots,
   {\delta_K\over \sigma_K}+Z_K\in J_K\Big\} $$
for $k=1,\ldots,K$, where $\delta_k=0$ for $k\ge k_0+1$, 
and where $Z_1,\ldots,Z_K$ are standard normals with correlations 
coming from $Q$. There is a corresponding result for $k_F$. 
This creates a different picture than with Ishwaran and Rao's Figure 1,
which has been produced under conditions corresponding to
having $|\delta_k|$ of infinite size for $k\le k_0$ 
(and $Q$ diagonal). 

Ishwaran and Rao comment that the model order parameter $k_0$
is not a smooth function of $\beta$, and as such falls outside
the standard regularity conditions used in our FMA and FIC papers.
The resulting predictors, say 
$\hatt\mu_B=x^\tr\hatt\beta_B$ and $\hatt\mu_F=x^\tr\hatt\beta_F$ 
for given covariate position $x$, are however amenable
to our methods, viewed as estimators of $\mu=x^\tr\beta$. 
The backwards and forwards predictors are model average methods
and can be analysed using the FMA methodology. 
Limit distributions are non-linear mixtures of biased normals,
and their performances may in particular be compared to that of 
AIC and FIC, as per Section 7 in the FMA paper. 
We also note that the arguments and results alluded to here 
should generalise without serious difficulties to e.g.~generalised 
linear models.  

Ishwaran and Rao `have always wondered about' whether it is better
to use forward or backward stepwise regression. They might perhaps
be encouraged to continue their fruitful wondering. Even in cases when 
$\hatt k_F$ is more successful than $\hatt k_B$ as an estimator 
of $k_0$ (where, as we argue above, the analysis and conclusion 
is less clear-cut than what it appears to be in their discussion), 
the backwards performance would still be better than forwards performance 
for predicting $x^\tr\beta$, 
in significant portions of the parameter space. 

We use this opportunity to nod in agreement to comments
made by Shen and Dough\-erty (Section 3), that it is very useful
when the list of candidate models can be restricted a priori,
for both FIC and FMA.  
In situations with a nested sequence of models, as above, 
this means reducing the number of candidates from $2^q$ to $q+1$.
On the other hand the list should be broad enough to reflect
real modelling information, as viewed in conjunction with
focus parameters. One possibility for shortening the queue of suitors
is via suitable thresholding and re-weighting, 
e.g.~including only the ten most promising models as monitored 
by the FIC scores, or by the posterior probabilities inside a BMA setup.
Our FMA theory continues to be applicable also for such strategies.  
 
\section 
\centerline{\bf 4. FMA versus BMA}

\hop
Raftery and Zheng come dressed as BMA's witnesses
and deliver a strong case. In their earnest zeal they
perhaps inadvertently risk classifying or portraying 
our FMA work as being anti-Bayesian, in spirit, intent or result. 
That would be a case of incorrect classification. 
Our FMA bag comprises not only the compromise estimators
of FMA's Section 4, but also averages of the generalised ridge estimators
developed in Section 8, and these again are close relatives
to BMA methods, as explained in Section 9. 
When developing our FMA methodology our points of motivation 
indeed included our wish to understand better 
the behaviour of BMA strategies.  

Realising that both `BMA' and `FMA' are big bags of methods, then, 
it is a little over-suggestive when Raftery and Zheng 
say that `BMA [was] generally found to have better performance' 
and that `FMA itself does not appear to yield optimal methods'. 
Some BMA regimes are better than others, and some FMA schemes have 
optimality properties. One may e.g.~work with model selection
schemes that post model selection use estimators that 
are minimax over say $\delta^\tr K^{-1}\delta\le c$ type regions,
using for this step methods similar to Blaker's (2000). 
Also, as mentioned above, some of the generalised ridge versions 
of FMA correspond (to the first order) to BMA schemes. 

Several model average schemes may be added to the annotated list
given in FMA's Sections 5 and 7. Ishwaran and Rao took up
backwards and forwards model selection procedures and as 
we explained above these may be analysed inside the FMA framework;
in particular, their behaviour may be analysed using FMA's Theorem 4.1. 
Johnson might consider having his former life prolongated by
revisiting his and his colleagues' robust Bayesian estimation 
methods, using the FMA apparatus to understand performance. 

Cook and Li discuss sliced inverse regression and 
central subspaces methods. Such methods are geared more 
towards dimension reduction than selection of subsets,
and may be compared to principal components regression
(see e.g.~Mardia, Kent and Bibby, 1979, Ch.~8)
and to partial least squares regression 
(see e.g.~Helland, 1990). With some work we believe versions 
of these dimension reduction methods may be characterised
and analysed as FMA methods. In these situations it would
be more natural to compare performances in terms of 
suitably averaged prediction accuracy, see our next section. 

As far as performance is concerned, Raftery and Zheng are perhaps right 
to arrest us for not paying enough attention to the existing BMA 
literature. They provide references to and give a summary of
three main strands of results: general Bayes theory (along with
studies of robustness to prior specifications); 
simulations; and cross-validation type predictive performance.
See also the concise and useful discussion in Clyde and George (2003).  
What we intended to point to in our introduction to the FMA paper
was the surprising lack in the literature 
of what one may think of as `the fourth strand of results',
namely limit distribution statements. 
In mathematical statistics we are not quite satisfied 
with simulations and cross validation and indications
of good performance; we need precise limit distribution results.
This is not only dictated by tradition and aesthetics, 
but gives practical mathematics, providing good approximations
for precision measures as well as a tool for comparing performances,
say of different BMA schemes. What is logistic regression 
without results about limiting behaviour of likelihood methods? 
What is years of hands-on experiences with averages 
without the central limit theorem? 

Shen and Dougherty stress, along with Johnson and with Raftery and Zheng, 
like we have done, the necessity of securing a well-defined 
interpretation of focus parameters (or variables) across models. 
In our framework this is taken care of via $\mu=\mu(f)$, where $f$ belongs 
to suitable submodels of the widest $f(y,\theta,\gamma)$ model. 
This requirement, when boomeranged back to BMA's watchtower,   
becomes the issue we raise in FMA's Section 1.1,
that BMA typically entails mixing together conflicting 
prior opinions about focus parameters. 
Our discussants do not take up this point. 

\section 
\centerline{\bf 5. Average quality of predictors}

\hop
We appreciate Raftery and Zheng's additional comments to
and extended analysis of the low birth weight data set. 
Our own analysis of this data set was primarily intended
as an illustration of the developed methods, as opposed
to a full scientific report on low birth weights. 
This was also why we chose somewhat simple parameters
as foci. Let us write these as $p_1=p(z_1)$, $p_2=p(z_2)$ 
and $\rho=p_2/p_1$, where $z_1$ and $z_2$ are the average covariate vectors 
for white and black mothers, respectively. We may agree
with Johnson and with Raftery and Zheng that there are yet other parameters
to focus on, with perhaps higher socio-biological relevance;
again, our parameters were chosen for illustration 
and simplicity. We still believe that $\rho$ has some
merit, though. A litmus test for `being of interest' might be
whether one can imagine a newspaper or magazine publishing a story 
about a finding concerning the parameter in question; 
here a news story sentence like 
`the average black mother has a 50\% greater chance 
than the average white mother of giving birth to 
too small children' would appear to pass the test.
Of course one should with such a finding attempt to 
investigate further, including aspects of the covariate
distributions.    

Comments from Cook and Li as well as from Raftery and Zheng
point to the usefulness of developing the FIC and FMA apparatus 
to assess prediction quality when averaged in suitable ways,
rather than for one focus parameter at a time. 
We touch on this in the FIC paper's Sections 5.6 and 7.2. 
Such averaging is particularly natural in regression models,
where focus might be on the behaviour of say $\hatt\mu(x,u)$
for a regression surface $\mu(x,u)$ for particular 
sub-regions of $u$ for fixed $x$, and so on. 
We note that the theory and arguments also invite 
suitable weighted generalisations of the AIC. 

To indicate how the machinery can be developed further, 
consider a linear regression setup with 
$Y_i=x_i^\tr\beta+u_i^\tr\gamma+\eps_i$ 
for $i=1,\ldots,n$, where the $\eps_i$s are i.i.d.~with mean zero 
and standard deviation $\sigma$, and where $\gamma=\delta/\rootn$. 
The $x_i$s are protected while elements of the $u_i$s 
may or may not be taken into the finally selected model. 
We study the average weighted prediction error 
$$\EE_n=n^{-1}\sumin(\hatt\xi_i-\xi_i)^2w(x_i,u_i), \eqno(5.1)$$
where $\xi_i=\E(Y\midd x_i,u_i)=x_i^\tr\beta+u_i^\tr\gamma$
and $\hatt\xi_i$ an estimator thereof, with
$w(x,u)$ a suitable weight function. 
We shall see that $n\EE_n$ has a limit distribution,
under reasonable conditions. 

Let  
$$\Sigma_n=n^{-1}\sumin\pmatrix{x_i \cr u_i \cr}
   \pmatrix{x_i \cr u_i \cr}^\tr
   =\pmatrix{\Sigma_{n,00} &\Sigma_{n,01} \cr
             \Sigma_{n,10} &\Sigma_{n,11} \cr}, $$
of size $(p+q)\times(p+q)$, assumed to be of full rank.
Its inverse $\Sigma_n^{-1}$ has blocks denoted $\Sigma_n^{ij}$,
and similarly for the smaller $(p+|S|)\times(p+|S|)$ matrix
$\Sigma_{n,S}$ with inverse $\Sigma_{n,S}^{-1}$. 
We assume that $\Sigma_n\arr\Sigma$ as $n$ increases,
also of full rank. We let $L_n=\Sigma_n^{11}$,
along with $L_{n,S}=(\pi_SL_n^{-1}\pi_S^\tr)^{-1}$ and
$H_{n,S}=L_n^{-1/2}\pi_S^\tr L_{n,S}\pi_SL_n^{-1/2}$. 
The matrices $L_n,L_{n,S},H_{n,S}$ have limits $L,L_S,H_S$. 

For the $S$ subset estimator
$$\pmatrix{\hatt\beta_S \cr \hatt\gamma_S \cr}
   =\Sigma_{n,S}^{-1}n^{-1}\sumin\pmatrix{x_i \cr u_{i,S}\cr}Y_i, $$
we have 
$$\rootn\pmatrix{\hatt\beta_S-\beta \cr \hatt\gamma_S \cr}
   \arr_d\pmatrix{C_S \cr D_S \cr}
   =\Sigma_S^{-1}
   \pmatrix{\Sigma_{01}\delta +M \cr \Sigma_{11}\delta + N_S\cr}, $$
where $(M,N)\sim\N_{p+q}(0,\sigma^2\Sigma)$ and $N_S=\pi_SN$. 
We may write 
$$\eqalign{
\EE_n&=n^{-1}\sumin(x_i^\tr\hatt\beta_S+u_{i,S}^\tr\hatt\gamma_S
   -x_i^\tr\beta-u_i^\tr\gamma)^2w(x_i,u_i) \cr
&=n^{-1}\sumin\Bigl[\pmatrix{x_i \cr u_i \cr}^\tr
   \pmatrix{\hatt\beta_S-\beta \cr 
   \pmatrix{\hatt\gamma_S-\gamma_S \cr -\gamma_{S^c} \cr} \cr} \Bigr]^2
   w(x_i,u_i) \cr
&=\pmatrix{\hatt\beta_S-\beta \cr \hatt\gamma_S - \gamma_S \cr
   -\gamma_{S^c} \cr}^\tr\Omega_n 
  \pmatrix{\hatt\beta_S-\beta \cr \hatt\gamma_S - \gamma_S \cr
   -\gamma_{S^c} \cr}, \cr} $$
where $\Omega_n$ is the $w$-weighted version of $\Sigma_n$ above.  
Thus, if only $\Omega_n\arr_p\Omega$,  
$$n\EE_n\arr_d \EE=
\pmatrix{C_S \cr D_S-\delta_S \cr -\delta_{S^c} \cr}^\tr
  \Omega 
\pmatrix{C_S \cr D_S-\delta_S \cr -\delta_{S^c}  \cr}. $$
Expressions for the mean of $\EE$ may be found using 
tools of the FIC paper. 

When $w=1$ in (5.1) we have $\Omega_n=\Sigma_n$
and a corresponding simplification for $\EE$. 
The limiting risk using $S$ can be shown to become 
$$\E(\EE)=(p+|S|)\sigma^2
   +\delta^\tr L^{-1/2}(I-H_S)L^{-1/2}\delta, $$  
using arguments as in FIC's Section 7.2. 
Let $D_n=\rootn\hatt\gamma_\full$, which goes to 
a $\N_q(\delta,\sigma^2 L)$. An unbiased risk estimator is 
$$\eqalign{
\hatt\risk_S
&=(p+|S|)\hatt\sigma^2+{\rm Tr}\bigl[L^{-1/2}(I-H_S)L^{-1/2}
   (D_nD_n^\tr-\hatt\sigma^2L)\bigr] \cr
&=(p-q+2|S|)\hatt\sigma^2+D_n^\tr L_n^{-1}D_n
   -D_n^\tr L_n^{-1/2}H_{n,S}L_n^{-1/2}D_n,\cr} $$
where $\hatt\sigma^2$ is the usual unbiased estimator of variance,
using the full model. 
This leads to the following selection criterion: 
choose the subset with smallest value of 
$$\hbox{ave-FIC}(S)
   =\hatt\sigma^2
   \{2|S|+n\hatt\phi^\tr (I-H_{n,S})\hatt\phi/\hatt\sigma^2\}, $$
where $\hatt\phi=L_n^{-1/2}\hatt\gamma_\full$. 
This appears to be related to both Mallows' $C_p$ 
as well as to Cook and Li's eq.~(2) (worked out there 
for the case of $p=1$, $x_i=1$, and $\sumin u_i=0$). 
Note that $\rootn\hatt\phi/\hatt\sigma\arr_d\N_q(\phi,I)$,
where $\phi=L^{-1/2}\delta$. For other extensions of 
the Mallows criterion, and theory, see Birg\'e and Massart (2001).  

We note that the above ideas and arguments may be used
to find precise limit distributions for 
average prediction error variables of the type
$\sumin\{\hatt\mu(x_i,u_i)-\mu(x_i,u_i)\}^2w(x_i,u_i)$,
in quite general regression models and for quite general
model average estimators. Such results may in particular
be used for model and subset selection purposes. 
One is quite free to choose weight schemes appropriate
for the purpose. If one wishes to assess predictor quality 
for a fixed $x_0$, when averaged over $u$, one may insert 
$w(x_0,u)$ proportional to an estimate of the conditional
density of $u$ given $x_0$. This might be a multinormal 
density, or a kernel-smooth over a window around $x_0$. 

We think that developments as above might lead to useful 
`focussed regression diagnostics' of different types.
The comments of Cook and Li also point in such directions. 

\section 
\centerline{\bf 6. Second order corrections}

\hop
In our papers we have determined the limit distribution
of $\Lambda_{n,S}=\rootn(\hatt\mu_S-\mu_\true)$ 
(as well as for more general estimators, like 
the post model selection estimator). This gives the approximation 
$$\risk_n(S,\delta)=n\,\E(\hatt\mu_S-\mu_\true)^2
   \doteq \E\Lambda_S^2, \eqno(6.1)$$ 
where $\Lambda_S$ is the limit variable. 
In FMA's Section 10.7 and FIC's Section 7.6 we mentioned
the potential for suitable finite-sample corrections
to the first order results of type (6.1).
We are glad that Tsai has taken up this challenge, providing 
what he terms `improved' and `corrected' versions of the FIC. 

The exact bias and variance of $\hatt\mu_S$ would often depend
in complicated ways on the model and sample size;
see e.g.~Duki\'c and Pe\~na (2003) for finite-sample
analysis of some particular post-selection estimators
in Gau{\ss}ian models.  
Sometimes expansions for these might be worked out, however. 
Suppose in general terms that 
$$\eqalign{
\E\Lambda_{n,S}&=B_1(S,\delta)
   +B_2(S,\delta)/\rootn+B_3(S,\delta)/n+o(1/n), \cr
\Var\,\Lambda_{n,S}&=V_1(S)+V_2(S,\delta)/n+o(1/n), \cr}$$
for suitable coefficients. Lemma 3.3 in the FMA paper gives
in fact expressions for the leading terms $B_1(S,\delta)$
and $V_1(S)$, and hence for the leading term 
$\E\Lambda_S^2=B_1(S,\delta)^2+V_1(S)$ in (6.1). 
In then follows that 
$$\eqalign{
{\rm risk}_n(S,\delta)
&=B_1(S,\delta)^2+V_1(S) 
   +2B_1(S,\delta)B_2(S,\delta)/\rootn \cr
&\qquad 
   +\{B_2(S,\delta)^2+2B_1(S,\delta)B_3(S,\delta)
   +V_2(S,\delta)\}/n+o(1/n). \cr}$$
This shows that the second order term to catch 
(and estimate) is $2B_1(S,\delta)B_2(S,\delta)/\rootn$. 
This necessitates finding an expression for $B_2(S,\delta)$. 

Following Tsai, this requires taking the delta method
one step further, using a 2nd order Taylor expansion. 
We do this in a somewhat different way. Starting with  
$$\hatt\mu_S-\mu_\true
 =\mu(\hatt\phi_S,\gamma_{0,S^c})-\mu(\phi_{0,S},\gamma_{0,S^c})
   +\mu(\theta_0,\gamma_0)-\mu(\theta_0,\gamma_0+\delta/\rootn), $$
we may split $\Lambda_{n,S}$ into two parts, with leading terms
$$(\dell\mu/\dell\phi)^\tr\rootn(\hatt\phi_S-\phi_{0,S})
   +\half\rootn(\hatt\phi_S-\phi_{0,S})^\tr
   \mu_{11,S}(\hatt\phi_S-\phi_{0,S}) $$
and 
$$-(\dell\mu/\dell\gamma)^\tr\delta
  -\half\delta^\tr\mu_{22}\delta/\rootn. $$
In our notation, $\mu_{11,S}$ is the $(p+|S|)\times(p+|S|)$ matrix
of 2nd order derivatives of $\mu(\theta,\gamma_S,\gamma_{0,S^c})$
w.r.t.~$(\theta_S,\gamma_S)$, 
while $\mu_{22}$ is the $q\times q$ matrix of 2nd order 
derivatives of $\mu(\theta,\gamma)$ w.r.t.~$\gamma$.
These derivatives are evaluated under the narrow model $(\theta_0,\gamma_0)$. 

To come further we need 
$$\E\rootn(\hatt\phi_S-\phi_{0,S})
  =J_S^{-1}\pmatrix{J_{01} \cr \pi_SJ_{11} \cr}\delta
   +m_S(\delta)/\rootn+n_S(\delta)/n+\cdots, $$
with suitable (but often cumbersome) expressions 
for $m_S(\delta)$ and $n_S(\delta)$
obtainable from work touched on by Tsai; see also 
Barndorff-Nielsen and Cox (1994, Chapters 5--6). 
This leads to  
$$B_2(S,\delta)=(\dell\mu/\dell\phi_S)^\tr m_S(\delta)
   +\half\Tr(\mu_{11,S}J_S^{-1})-\half\delta^\tr\mu_{22}\delta. $$
To summarise, 
$$\risk_n(S,\delta)=\E\Lambda_S^2
   +2B_1(S,\delta)B_2(S,\delta)/\rootn+o(1/\rootn) \eqno(6.2)$$
provides a 2nd order corrected version of (6.1). 

The treatment above is related to but not fully equivalent to 
what Tsai does. He studies nonlinearity aspects in his Section 2
and bias of likelihood estimators in Section 3. 
It appears to us, from the arguments above, 
that it is necessary to combine both these 2nd order aspects. 
If not one risks catching one or two of the terms making up 
$B_2(S,\delta)$, but not all three, and a partial reparation
might be worse than no reparation. 

We would perhaps hesitate to affix the labels `improved' and `corrected'
too firmly on Tsai's modified FICs. It is clear from the above
that there are several possibilities for such 2nd order 
approximations to the mean squared error of estimators.
Also, one needs indirectly or directly to estimate 
$B_1(S,\delta)B_2(S,\delta)$ from data, where there are 
several paths to follow, e.g.~regarding wide versus narrow 
estimation of partial derivatives. Furthermore, this
estimation step might cause additional variability 
that might take away the intended benefit.
Such phenomena are well-known in mathematical statistics. A 2nd order 
Edgeworth expansion might not be a genuine improvement 
over a 1st order Edgeworth expansion, for example, or perhaps
there is improvement only for very large sample sizes. 
All this serves to indicate that further studies are required
before a general-purpose 2nd order FIC can be established. 

We note that Tsai's work, and presumably also the development above, 
is relevant also when it comes to assessing behaviour of 
model average estimators. 

\section
\centerline{\bf 7. Estimators from other likelihoods}

\hop
Tsai points out that $\gamma$ parameters sometimes are in focus,
and we agree. Our FIC and FMA apparatus is nicely able to handle
this, since it covers all smooth $\mu(\theta,\gamma)$ parameters;
Tsai appears to claim otherwise. With focus on $\gamma_j$
we find $\diff=-e_j$, with $e_j=(0,\ldots,1,\ldots,0)^\tr$
being the $j$th unit vector. We are free to form general model
average estimators $\hatt\gamma_j=\sum_S c(S\midd D_n)\hatt\gamma_{j,S}$,
where incidentally terms with $S$ not touching $j$ will be equal to zero. 
Using FMA's Theorem 4.1, we find 
$$\rootn\{\hatt\gamma_j-(\gamma_{0,j}+\delta_j/\rootn)\}
  \arr_d \hatt\delta_j(D)-\delta_j 
  \quad {\rm for\ }j=1,\ldots,q, $$
and so on. The FIC can also be applied, and one may study
simultaneous estimation of the full $\gamma$ vector. 
It is also natural to include the goodness-of-fit measure
$$\hatt\delta^\tr\hatt K^{-1}\hatt\delta
   =n(\hatt\gamma-\gamma_0)^\tr\hatt K^{-1}(\hatt\gamma-\gamma_0) $$
in the data analysis. It is a $\chi^2_q(\delta^\tr K^{-1}\delta)$
in the limit.

There might be situations where the ordinary likelihood
apparatus cannot be used, or can be expected to perform
poorly, and where variations like profile likelihoods,
empirical likelihoods and quasi-likelihoods may be helpful. 
This would call for extensions of our work. 
We do not think, however, that profiling is necessary, 
or that it leads to new results,
inside our parametric $f(y,\theta,\gamma)$ framework.
We are therefore puzzled with Tsai's elaborations in this regard;
under weak conditions the $S$-model profile likelihood estimator
of $\mu$ will simply be our old maximum likelihood estimator $\hatt\mu_S$.
Tsai's intricate definition of a new `random parameter' $\mu_{\rm prof,true}$ 
does not correspond to our more naturally defined $\mu_\true$. 

In Hjort and Claeskens (2003) we report on extensions 
of our FIC and FMA work, for model selection and model averaging 
inside the semiparametric Cox regression model. Focus parameters could
take the form $\mu(\beta,H,z)$, involving the parametric as well
as the nonparametric part of the model, as with the median time
to survival for a patient with given covariates $z$. Our existing
theory will be seen to go through without essential modifications
as long as $\mu$ is a function of $\beta$ and covariates only,
whereas such modifications are called for when it also involves $H$. 
Similarly, extensions may be envisaged for use in spatial models
with covariates, inside particular formats of parameter estimation;
we in particular have in mind the pseudo-likelihood method
of Besag (1974, 1977) for Markov random fields;
the quasi-likelihood of Hjort and Omre (1994, Section 3) 
for spatial correlation models; 
and various methods for observed and aggregated point processes
reviewed in Richardson (2003). 

\section 
\centerline{\bf 8. Supplementary comments}

\subsection
{\sl 8.1. When $p$ and $q$ become big.} 
Our methodology has been developed under the classic 
asymptotics scenario where the number of parameters stays bounded 
when the sample size increases. Shen and Dougherty point out 
(in their Section 4) that the results might need modifications
to apply when $p+q$ is big, as will happen in many 
potential applications. We agree. This needs further mathematical
developments. We do believe, however, that our asymptotics results 
will continue to provide adequate descriptions and 
approximations even when $p+q$ grows with $n$, 
but slowly enough to have $p+q=o(\rootn)$. Establishing
such results would need further work, but might use 
methods similar to those used in e.g.~Portnoy (1988). 

We use the opportunity to opine that if $p+q$ becomes too big,
it should be reduced. If one has 1,000 covariates per patient,
one does good to compress and synthesise these, using 
substantive prior knowledge along with statistical techniques,
before throwing the data set to a regression selector or averager. 
Also, methods like principal components and partial least squares
regression might easily perform better than subset finding schemes. 

\subsection
{\sl 8.2. Loss functions and aspects of costs.}
Cook and Li point out that using limiting mean squared error
will not always suffice for making the relevant 
conclusions, regarding e.g.~model selection;
see also comments by Shen and Dougherty.  
In some cases there is a cost $k(S)$ associated with
observing future data for regressors in index set $S$. 
With loss functions that suitably combine precision
with cost, like $n(\hatt\mu-\mu)^2+\alpha k(S)$, we would have 
$$\E\,{\rm loss}_n(S)=n\,\E(\hatt\mu_S-\mu_\true)^2+\alpha k(S)
   \arr \E\Lambda_S^2+\alpha k(S). $$
This might be estimated using a slight extension of the FIC,
after which an optimal subset may be extracted. 

We have favoured limiting mean squared error as performance
criterion, but might also have worked e.g.~with $L_1$ loss,
leading however to more complicated expressions for 
and estimators for $\E|\Lambda_S|$ and so on. 

\subsection
{\sl 8.3. Handling corner parameters.}
Shen and Dougherty discuss a general four-para\-meter model where 
rate measurements are of the form 
$V(x_1,x_2\midd\beta_1,\beta_2,\beta_3,\beta_4)$
plus observation error, with 
$$V={\beta_1x_1\over \beta_2(1+\beta_3x_2)+x_1(1+\beta_4x_2)}. $$
The case of $(\beta_3,\beta_4)=(0,0)$ is the so-called
Victor--Michaelis--Menten model for enzyme mediated reactions.
In fisheries research it is also well known as the spawner--recruit model, 
dating back to an influential paper of Beverton and Holt (1957); 
see Gavaris and Ianelli (2002) and 
the engaging discussion in Smith (1994, Ch.~8).
Shen and Dougherty discuss aspects of modelling the $V$, 
in particular looking at the four possibilities 
in-in, in-out, out-in, out-out for $(\beta_3,\beta_4)$. 
This cannot be studied well without a clearer understanding 
of the error structure involved. That this is non-trivial and vital,
and will vary widely with context, is clear from 
Ruppert, Cressie and Carroll (1989).
Shen and Dougherty allude to pre-test methods, which decide 
on in- or exclusion of $\beta_3$ and $\beta_4$ on the basis of 
tests for their presence. We note that such schemes are 
again model average methods, and fall inside our developed theory. 

There might sometimes be situations where it is known a priori
that e.g.~$\beta_3\ge0$, $\beta_4\ge0$. The theory we have developed
presupposes that $(\beta_3,\beta_4)$ is an inner point of the
parameter space. To handle `corner problems', like here, 
one needs somewhat more intricate methods, which would depend
more on the specifics of the problem. See Hjort (1994b) 
for one such example, concerned with compromise estimators 
when the $t$ family is used as an extension of the normal
in e.g.~regression settings. Similar problems emerge 
in models with variance components. Methods of Vu and Zhou (1997)
appear relevant when attempting to generalise our results
to corner parameters.   

An opinion perhaps too rarely expressed, which we share, 
is that statisticians should be more eager to help develop good
non-linear regression models, as here. The comfort and ease
with which we reach moderately adequate approximations 
and inference precision using the flexible machinery
of (generalised) linear models may sometimes take the edge out of our
professional modelling creativity. 

\subsection 
{\sl 8.4. Non-nested models.}
We have for the most part stayed inside a framework where
the biggest model is thought to be correct. 
Cook and Li mention the problem of non-nested models. 
The simplest answer, perhaps, from a principled point of view,
is that one might search for a bigger model formulation that
encompasses both. Consider estimating the median, for example,
including under view both the gamma and the log-normal models. 
One may then work with estimators of the type 
$\hatt\mu=W\hatt\mu_{\rm gam}+(1-W)\hatt\mu_{\rm logn}$, 
with weights somehow dictated by data, e.g.~via goodness-of-fit
measures, or via closeness of the two estimates involved 
to the nonparametric $\hatt\mu_{\rm nonpm}$, i.e.~the sample median.
Behaviour and performance may be studied using our methods. 

There are examples in science where non-nested and somehow 
conflicting statistical theories are not easily resolved,
of course. A controversy of some fame inside fisheries research, 
and that has perhaps not yet been solved to satisfaction 
despite having been pondered over for about a hundred years, 
is the Dannevig vs.~Hjort case. 
It is concerned with models for spawning, recruitment, migration
and development of fish populations. 
Dannevig essentially believed 
in a deterministic relationship between the number of recruits 
and the number of yolk-sack codfish larvae, whereas Hjort 
argued that it is the environmental conditions during the critical
phases of development that play the more important roles. 
He was able to develop year-class assessment methods, 
collect relevant data and utilise actuarial mathematical
methods of the time to substantiate and refine his theories; 
cf.~e.g.~Hjort (1914). 
See {\tt http://www.math.ntnu.no/{$\sim$}ingeol/bemata/}, 
where a study of structured stochastic models has been launched,
involving computer-intensive inference in biological marine systems, 
and the interesting discussion in Smith (1994) and Secor (2002). 
This may be an example where model averaging might be useful, 
in a non-nested setup, mixing predictions of e.g.~next season's 
abundance (perhaps as a function of quota thresholds) 
using elements of both scientific models. 

\subsection 
{\sl 8.5. Interpreting FIC numbers.}
The FIC scores have been developed as estimates of 
$n$ times mean squared error of subset estimators 
(modulo an additive constant), and as such depend on the scale used.
They may be made scale-independent via say 
$$\FIC^*(S)=\hatt\FIC(S)/\hatt\diff^\tr\hatt K\hatt\diff, $$
as in FMA's Section 5.3. This would make comparison 
and interpretation easier across applications. 
We would in particular have 
$$\FIC^*(\full)=2 
  \quad {\rm and} \quad 
  \FIC^*({\rm narrow})=n\{\hatt\diff^\tr(\hatt\gamma_\full-\gamma_0)\}^2
   /\hatt\diff^\tr\hatt K\hatt\diff. $$  

\subsection
{\sl 8.6. When is $\diff$ equal to zero?}
We have seen that the behaviour of model average estimators
is critically determined by $\diff=J_{10}J_{00}^{-1}\dellone-\delltwo$.
In particular, if $\diff=0$, then all subset and model average
estimators are asymptotically equivalent to the narrow model estimator;
$\rootn(\hatt\mu-\mu_\true)\arr_d\N(0,\tau_0^2)$ for all reasonable
competitors. 
The typical situation leading to $\diff=0$ is when the parameter
does not depend on $\gamma$ and in addition $\theta$ and $\gamma$
are orthogonal parameters, in the sense that their full model
estimators are independent in the limit, i.e.~$J_{01}=0$.
Johnson asks whether $\diff$ may be zero also in other situations. 
Here is one example, in the framework of the 
exponential-within-Weibull model of FMA's Section 4.4. 
Assume we wish to estimate the $\alpha$-quantile
$\mu=\nu^{1/\gamma}/\theta$, where $\nu=-\log(1-\alpha)$. 
Then calculations give $\diff=(\nu/\theta)\,\{-(1-r)+\log\nu\}$. 
For estimating the $\alpha=0.7826$-quantile, therefore, 
$\diff$ happens to be equal to zero. 

\bigskip
\centerline{\bf Additional references} 

\def\ref#1{{\noindent\hangafter=1\hangindent=20pt
  #1\smallskip}}          
\parindent0pt
\baselineskip11pt
\parskip3pt 

\medskip
\smallskip

\ref{%
Barndorff-Nielsen, O.E.~and Cox, D.R. (1994).
{\sl Inference and Asymptotics.}
Chapman \& Hall, London.}

\ref{%
Besag, J. (1974).
Spatial interaction and the statistical analysis of lattive systems
[with discussion]. 
{\sl Journal of the Royal Statistical Society} {\bf B 36}, 195--225.}

\ref{%
Besag, J. (1977).
Efficiency of pseudolikelihood estimation for simple Gaussian fields.
{\sl Biometrika} {\bf 82}, 616--618.} 

\ref{%
Beverton, R.J.H.~and Holt, S.J. (1957).
On the dynamics of exploited fish populations.
Ministry of Agriculture, Fisheries and Food (UK),
{\sl Fisheries Investigations} (Series 2) {\bf 19}.}

\ref{%
Birg\'e, L.~and Massart, P. (2001).
Gaussian model selection.
{\sl Journal of the European Mathematical Society} {\bf 3}, 203--268.}

\ref{%
Blaker, H. (2000).
Minimax estimation in linear regression under restrictions.
{\sl Journal of Statistical Planning and Inference} {\bf 90}, 35--55.}

\ref{%
Brown, L.D. (2000).
An essay on statistical decision theory. 
{\sl Journal of the American Statistical Association} {\bf 95}, 1277--1281.
Also in {\sl Statistics in the 21st Century}
(2002, eds.~A.E.~Raftery, M.A.~Tanner and M.T.~Wells),
Chapman \& Hall/CRC, London.} 

\ref{%
Claeskens, G.~and Hjort, N.L. (2003). 
Goodness of fit via nonparametric likelihood ratios.
Submitted for publication.} 

\ref{%
Clyde, M.~and George, E. (2003).
Model uncertainty. 
ISDS Technical Report 03--17, Duke University.}

\ref{%
Duki\'c, V.D.~and Pe\~na, E.A. (2003).
Estimation after model selection in a Gaussian model.
{\sl Journal of the American Statistical Association}, to appear.}

\ref{%
Gavaris, S.~and Ianelli, S.N. (2002).
Statistical issues in fisheries' stock assessments [with discussion].
{\sl Scandinavian Journal of Statistics} {\bf 29}, 245--271.}

\ref{%
Helland, I.
Partial least squares regression and statistical models.
{\sl Scandinavian Journal of Statistics} {\bf 17}, 97--114.}

\ref{%
Hjort, J. (1914).
Fluctuations in the great fisheries of Northern Europe.
{\sl Rapports et Proc\`es-Verbaux des R\'eunions
du Conseil International pour l'Exploration de la Mer}
{\bf 20}, 1--228.}

\ref{%
Hjort, N.L. (1994a). 
Should the Olympic sprint skaters run the 500 meter twice?
Statistical Research Report, 
Department of Mathematics, University of Oslo.}

\ref{%
Hjort, N.L. (1994b). 
The exact amount of t-ness that the normal model can tolerate.
{\sl Journal of the American Statistical Association} {\bf 89}, 665--675.}

\ref{%
Hjort, N.L.~and Omre, H. (1994). 
Topics in spatial statistics [with discussions].
{\sl Scandinavian Journal of Statistics} {\bf 21}, 289--357.}

\ref{%
Hjort, N.L.~and Rosa, D. (1999).
Who won? {\sl Speedskating World} {\bf 4}, No.~8, 15--18.}

\ref{%
Hjort, N.L.~and Claeskens, G. (2003).
Model averaging and focussed model selection
for Cox regression. Manuscript.}

\ref{%
Mardia, K., Kent, J.T.~and Bibby, J.M. (1979).
{\sl Multivariate Analysis.}
Academic Press, London.}

\ref{%
McCullagh, P. (2002). 
What is a statistical model? [with discussion]
{\sl Annals of Statistics} {\bf 30}, 1225--1308.}

\ref{%
Portnoy, S. (1988).
Asymptotic behavior of likelihood methods for exponential 
families when the number of parameters tends to infinity. 
{\sl Annals of Statistics} {\bf 16}, 356--366.} 

\ref{%
Richardson, S. (2003).
Spatial models in epidemiological applications [with discussion].
In {\sl Highly Structured Stochastic Systems}
(eds.~P.J.~Green, N.L.~Hjort and S.~Richardson).
Oxford University Press, London, pp.~237--269.}

\ref{%
Secor, D.H. (2002).
Historical roots of the migration triangle.
{\sl ICES Marine Science Symposia} {\bf 215}, 329--335.}

\ref{%
Smith, T. (1994). 
{\sl Scaling Fisheries:
The Science of Measuring the Effects of Fishing, 
1855--1955.}
Cambridge University Press, Cambridge.}

\ref{%
van der Vaart, A. (1998).
{\sl Asymptotic Statistics.}
Cambridge University Press.}

\ref{%
Vu, H.T.V.~and Zhou, S. (1997). 
Generalization of likelihood ratio tests under
nonstandard conditions.  
{\sl Annals of Statistics} {\bf 25}, 897--916.}

\bye